\documentclass[aps,pre,twocolumn,floatfix,showpacs]{revtex4}
\usepackage{graphicx}
\usepackage{bm}
\usepackage{amsmath}
\usepackage{subfigure}
\bibstyle{apsrev.bst}

\begin{document}
\title{Modelling background intensity in Affymetrix Genechips}
\author{K.M. Kroll}
\affiliation{Interdisciplinary Research Institute, Cit\'e
Scientifique BP 60069, F-59652 Villeneuve d'Ascq, France}
\affiliation{Institute for Theoretical Physics, Katholieke Universiteit
Leuven, Celestijnenlaan 200D, B-3000 Leuven, Belgium}
\author{G.T. Barkema}
\affiliation{Institute for Theoretical Physics, University of Utrecht, 
Leuvenlaan 4, 3584 CE, Utrecht, The Netherlands}
\affiliation{Institute-Lorentz for Theoretical Physics, University of Leiden, 
Niels Bohrweg 2, 2333 CA Leiden, The Netherlands}
\author{E. Carlon}
\affiliation{Institute for Theoretical Physics, Katholieke Universiteit
Leuven, Celestijnenlaan 200D, B-3000 Leuven, Belgium}
\date{\today}

\begin{abstract}
DNA microarrays are devices that are able, in principle, to detect
and quantify the presence of specific nucleic acid sequences in
complex biological mixtures. The measurement consists in detecting
fluorescence signals from several spots on the microarray surface
onto which different probe sequences are grafted. One of the problems
of the data analysis is that the signal contains a noisy background
component due to non-specific binding. This paper presents a physical
model for background estimation in Affymetrix Genechips. It combines two
different approaches.  The first is based on the sequence composition,
specifically its sequence dependent hybridization affinity.  The second
is based on the strong correlation of intensities from locations which
are the physical neighbors of a specific spot on the chip. Both effects
are incorporated in a background functional which contains $24$ free
parameters, fixed by minimization on a training data set. In all data
analyzed the sequence specific parameters, obtained by minimization, are
found to strongly correlate with empirically determined stacking free
energies for RNA/DNA hybridization in solution. Moreover, there is an
overall agreement with experimental background data and we show that the
physics-based model proposed in this paper performs on average better
than purely statistical approaches for background calculations. The
model thus provides an interesting alternative method for background
subtraction schemes in Affymetrix Genechips.
\end{abstract}


\maketitle

\newcommand{\ul}{\underline}
\newcommand{\bc}{\begin{center}}
\newcommand{\ec}{\end{center}}
\newcommand{\be}{\begin{equation}}
\newcommand{\ee}{\end{equation}}
\newcommand{\ba}{\begin{array}}
\newcommand{\ea}{\end{array}}
\newcommand{\beqn}{\begin{eqnarray}}
\newcommand{\eeqn}{\end{eqnarray}}

\section{Introduction}
\label{sec:intro}

DNA microarrays have become a powerful tool to monitor the gene
expression level of thousands of genes simultaneously on a genome-wide
scale (for a recent review see for instance Ref.~\cite{stoug05}).
Microarrays are based on the hybridization between the surface-bound DNA
sequences (called probes) and DNA or RNA sequences in solution (called
targets). The probes are designed to have a sequence exactly complementary
to that of the desired target sequence one wishes to detect in solution.
As the target molecules in solution are labelled with fluorescent markers,
the amount of hybridized targets can be determined by means of optical
measurements. The fluorescence intensity measured at a specific spot
on the microarray reflects the concentration of complementary targets
in the used sample solution.

One of the most prominent commercial platforms of DNA microarrays
is provided by Affymetrix~\cite{lips99}. By virtue of in-situ
photolithographic techniques Affymetrix produces arrays in which more than
one million different probes are grafted on a single chip. The probes
are 25 nucleotides long sequences of single-stranded DNA.  As a single
25-mer may not provide reliable measurements of the expression level
of one specific gene, Affymetrix chooses 10-16 fragments of different
regions for each gene, which together form a so-called probe set. Each
probe set is to uniquely characterize a given gene.

One of the problems of the data analysis is that the measured fluorescence
signal does not only contain information about the concentration of a
specific gene in solution, but also of other sources of hybridization
with fragments which only partially overlap with the surface-bound
sequence. Thus, the measured fluorescence of a given probe site can be
written as
\be 
I = I_0 + I_{sp} (c) 
\label{def_backg} 
\ee 
where $I_{sp} (c)$ is the specific contribution of the signal which
depends on the concentration $c$ of the complementary target in solution
and $I_0$ is a background signal. The aim of this work is to introduce
a new model which is based upon inputs from physical chemistry for the
calculation of $I_0$ for Affymetrix arrays.  Identifying the main sources
of background intensity is crucial in order to make accurate and reliable
estimates of gene expression levels mainly for weakly expressed genes,
for which $I_{sp} (c) \approx I_0$.

A peculiarity of Affymetrix Genechips is that probes come in pairs:
a probe, the so-called perfect match (PM), has a sequence exactly
complementary to the sequence in solution. A second probe, physically
located as neighbor of the PM in the chip, has a single non-complementary
base with respect to the specific target. The latter is known as mismatch
(MM).  Originally, MM's were supposed to estimate only the non-specific
hybridization, i.e.~it was expected that $I_{MM} \approx  I_0$, so that
from eq.~(\ref{def_backg}) one could have estimated $I_{sp} (c) = I_{PM} -
I_{MM}$. However, this approach experiences some difficulties as in some chips
as many as $30\%$ of the MM intensities are higher than the corresponding
PM's~\cite{naef03} (although this seems to occur predominantly in
low intensities regimes, where both PM and MM signals may be dominated
by non-specific hybridization \cite{bind05}). Moreover, it has been found
that $I_{MM}$ also depends on the concentration in solution of the almost
complementary target sequence.  Hence the background adjustment based
on the difference $I_{PM} - I_{MM}$ currently does not receive much
consensus and other strategies have been devised ~\cite{gent03}. For a
discussion of MM hybridization see also Refs.~\cite{carl06b,naef06,ferr07}.

Due to its central importance, the modeling of background intensities
is not new. One can distinguish here between models using purely
statistical treatment ~\cite{iriz03_sh,affy_statalgo,affy_mas5,liwo01}
and others where physical inputs coming from equilibrium thermodynamics
were employed~\cite{zhan03,held03,heks03,carl06,iriz03}. A more extensive
discussion of previous studies in relation with our results is postponed
to the final section of this paper.

In this paper we present a new method to estimate the background noise
of Affymetrix gene expression arrays. We construct a functional which
contains $24$ parameters, fixed by minimization on a set of training data.
The functional takes into account the physical chemistry of hybridization
by a subset of the $24$ parameters. These parameters depend on sequence
composition and which are equivalent to the stacking free energies in
the nearest-neigbor model~\cite{bloo00}. We also exploit the observation
that the background signal of a given site strongly correlates with the
intensities measured on neighboring sites. The accuracy of the results
is tested on a set of spike-in data in which transcripts are added in
solution at known concentration. In particular, being interested in
the accuracy of our background predictions, we focus on the data at
zero concentration.  The model developed in this paper reproduces the
spike-in data very well and in this particular case it performs better
than other popular algorithms used for background adjustment in Affymetrix
expression chips.

This paper is organized as follows: the background functional is
introduced in Sec.~\ref{sec:model}. The results of the minimization
are given in Sec.~\ref{sec:results}, where they are tested on the
spike-in data set and compared with the prediction of other algorithms.
Finally in Sec.~\ref{sec:discussion} we present a general discussion of
the results obtained and provide some general conclusions.

\section{Model}
\label{sec:model}

\begin{figure}[h!]
\includegraphics[width=8cm]{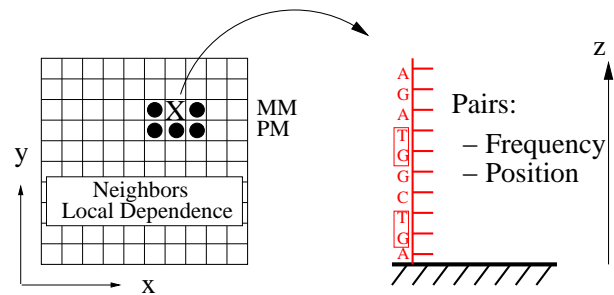}
\caption{Schematic view of the two main ingredients for the background
functional developed in this paper. (Left) Background intensity is
correlated to the fluorescence signal measured from neighboring spots.
(Right) The background depends also on the sequence dinucleotide
composition and on the relative distance of the dinucleotides from the
surface, i.e.~inhomogeneities are taken into account.}
\label{fig:orig_param}
\end{figure}

Our approach to estimate the background intensity is twofold. First,
we make use of the property of Affymetrix microarrays that neighboring
probes have similar sequences, and hence also similar affinities for
non-specific binding.  We recall that a fluorescence image from an
Affymetrix chip is contained in a file giving the $(x,y)$ coordinate of
the probe and the corresponding measured intensity. By setup a PM probe
is located at $(x,y)$, with odd $y$, and the corresponding MM probe is
located at $(x,y+1)$.  Hence the chip is arranged in rows of PM and MM
sequences, as shown in Fig.~\ref{fig:orig_param}. PM and MM pair probes
share all nucleotides but the middle (13$^{th}$) one. Hence there is a
strong sequence correlation between the rows with odd $y$ and the rows
at $y+1$.  But the sequences of neighbors along the $x$-direction are
also correlated, as part of the microarray design.

The main idea of our approach is to use MM intensities as background
estimates for the PM signals only for genes that are sufficiently low
expressed, i.e. for which both PM and MM signals are low on the global
scale of intensities in Affymetrix chips. An estimator is built up and
optimized around these low-intensity data (to be defined more precisely
later), which can then be applied to the whole chip, thus also in the
high intensity regimes. Let us consider a MM at position $(x,y)$. Because
of correlations with the neighboring sequences its intensity value will
be correlated to the intensities of the neighboring sites in the
chip. In particular, we consider the weighted average of intensities of
the two neighboring MM's at positions $(x \pm 1,y)$, of the corresponding
PM $(x,y-1)$ and of the two PM's at positions $(x \pm 1,y-1)$, as shown
in Fig.~\ref{fig:orig_param}.  Differences in sequences tend to cause
gaussian fluctuations in the effective affinities, hence in the logarithm
of the intensities $\eta(x,y)=\ln I(x,y)$, rather than in the intensities.
The local dependence of the background functional takes thus the form
\begin{eqnarray}\label{eq:eta_local}
\eta_{local}(x,y) &=& p_0 + p_1 \, \eta(x+1,y)
+ p_2 \, \eta(x-1,y) \nonumber\\
&+& p_3 \, \eta(x+1,y-1)+ p_4 \, \eta(x-1,y-1)\nonumber\\ 
&+&p_5 \, \eta(x,y-1),
\end{eqnarray}
in which $p_i, i=0\dots 5$ are weight factors, constrained by
$\sum_{i=0}^5 p_i=1$.

A completely different indicator of the background intensity is
purely based on the probe sequence. A well-known model to estimate the
affinity between a DNA probe and its complementary RNA target is the
nearest-neighbor model~\cite{bloo00}. Here, the affinity is given by
a summation over pairs of neighboring nucleotides, in which each term
can take 16 different values, depending on whether the sequence is AA,
AC, AG, AT, $\dots$, GG. We expect that the background signal is due to
the binding to the probe sequence of short fragments of sequences from
other genes, which are complementary to the probe over some fraction
of its total length.  We introduce 16 pair-strengths $p_{\alpha\beta}$
(with $\alpha, \beta \in \{{\rm A,T,C,G} \}$) as fitting parameters.

To approximately incorporate the effects of ``unzipping" of the DNA/RNA
hybrid on its top and bottom, we add a parabolic weighting as a function
of the position along the probe, around the middle of the probe at
$k_m=12\frac{1}{2}$. In the fabrication of the chip the majority of
probes does not reach its full length of $25$ nucleotides. The effect
of length variation is modeled by linear deviations in this weighting
function as well. In total, this yields
\begin{widetext}
\begin{equation}\label{eq:eta_seq} 
\eta_{seq}(s) = \sum_{k=1}^{24}
  \sum_{\alpha,\beta} \delta_{\alpha\beta}^{k}(s)\,p_{\alpha\beta}\,
  [1 + (k-k_m) \, p_{l}+ (k-k_m)^2\, p_{p}]
\end{equation}
\end{widetext}
with $\alpha, \beta \in \{\text{A,T,C,G}\}$ and where
\begin{equation}\label{eq:delta_alphabeta}
 \delta_{\alpha\beta}^{k}(s) =
\begin{cases}
 1 & \text{if}\,s_{k}=\alpha\,\mbox{and}\,s_{k+1}=\beta\\
 0 &\text{otherwise.}
\end{cases}
\end{equation}

Here $s_k$ indicates the $k$-th oligonucleotide of the sequence $s(x,y)$
of a total length of 25 letters. Summing over all possible letters
$\alpha, \beta$ is equivalent to counting the frequency of each pair
$\alpha\beta$ within a given sequence $s(x,y)$. The 16 parameters
$p_{\alpha\beta}$ reflect the influence of each pair $\alpha\beta$
on the background intensity. According to the nearest-neighbor model,
the parameters can be used to describe the formation of RNA/DNA hybrid
duplexes~\cite{sugi95_sh}.  Also here, we expect that our approximations
lead to a more or less gaussian spreading in the effective affinities. The
sequence-based estimation of the background intensity is then given by
$I_{seq}(x,y)=\exp (\eta_{seq} (s))$.

We then combine the two different estimates for the background affinity
with arbitrary weights:
\begin{equation}\label{eq:eta}
\ln I(x,y;s) \equiv \eta(x,y;s) = \eta_{local}(x,y) + \eta_{seq}(s)\\
\end{equation}
where the relative weight for the first estimate is absorbed in the
parameters $p_i$ --- we no longer inforce the restriction $\sum_{i=0}^5
p_i=1$ --- and the relative weight for the second estimate in the
parameters $p_{\alpha\beta}$.

We proceed by constructing a cost function whose minimization allows to
obtain estimates of the 24 parameters in Eq.(~\ref{eq:eta}).
We write the cost function as an average over all probes of the
squared difference between the actual background affinity and the prediction
$\eta(x,y;s')$:

\begin{equation}\label{eq:costfct}
\mathcal{S} = \frac{1}{N}\,\sum_{s'(x,y)} [\log{I_{MM}}-\eta(x,y;s')]^{2}.
\end{equation}
Here, $s'(x,y)$ is a subset of $N$ sequences which includes only
sequences of those MM intensities whose corresponding PM intensities
are below a certain threshold $\hat{\imath}$ (in Affymetrix units) and
which themselves do not exceed $\hat{\imath}$ to exclude bright MMs from
the analysis~\cite{naef03}.

Equation~(\ref{eq:costfct}) incorporates Affymetrix' original idea of
using MM intensities as background measures. Strict selection rules need
to be imposed on the input data ($s'(x,y)$) to ensure that only those
experimentally obtained values of $I_{MM}$ are used which can be clearly
attributed to background noise, and not to hybridization with the target
complementary to the corresponding PM probe. But how do we find a criteria
which identifies and filters the undesired probes? Fortunately, the
Spike-In data at concentration $c=0$ (for details, see~\ref{subsec:ed})
can be used as reference for background noise. By comparing the $I_{MM}$
histograms of the input and Spike-In data ($c=0$), a threshold intensity
$\hat{\imath}$ can be found such that both histograms are strongly
correlated. In the present work, the threshold intensity is set to
$\hat{\imath}=350$ resulting in a discard of $25-30\%$ of the data. For
comparison, saturated probes have intensities around 12,000.

The optimization algorithm used to perform the minimization of the
cost function given in Eq.~(\ref{eq:costfct}) is steepest descent
with damped newtonian dynamics, in which Eq.~(\ref{eq:costfct})
is interpreted as potential energy. The value obtained from the
minimization procedure can be used as a measure for the quality of the
attained minimum. 

\section{Results}
\label{sec:results}

\subsection{Experimental data}
\label{subsec:ed}

In the present work, we analyze data from Affymetrix
microarray experiments which are publicly available
under~\cite{latinsq,geo}. The results (scanned intensities) of each
experiment are saved in a so-called ``CEL-file'' (extension .CEL). For
each probe, the CEL-file contains information about its physical location
on the chip ($x$- and $y$-coordinate) and the mean intensity.
The CEL-file
does not contain any information about the probe set name or sequence.
For further processing the so-called CDF-file (chip description
file) is needed. Each CEL-file is associated with a CDF-file which
allows to retrieve the information necessary to map each probe to its
corresponding probe set. The sequence information can be found in the
probe-tag-file. The probe-tag-file contains the name of each probe,
its location, an Affymetrix specific probe interrogation position, the
sequence and the target strandedness. The latter file is particularily
useful if one wishes to investigate the sequence dependence of the
measured intensities.

Affymetrix offers a large palette of gene expression arrays for different
organisms. In this work, we focus on the analysis of two human genome
chipsets, namely HGU95A and HGU133A, and two non-human organisms, the
african clawed frog (Xenopus Laevis) and the zebrafish (Danio Rerio).
All four chipsets are used, in a first step, to investigate and validate
the correlation between well-known hybridization stacking energies and the
16 parameters of Eq.~(\ref{eq:eta}) (see Section~\ref{subsec:sdp}).  As
second step, we focus our attention to a subset of the HGU95A and HGU133A
datasets, the so-called Latin Square Experiments \cite{latinsq}. Those
experiments serve as calibration experiments as some target sequences
are added at controlled concentrations (``spiked-in'') to a background
reference solution. The target concentrations range from 0~pM to 1024~pM.
Since the spike-in experiments at zero concentration measure pure
background, we use them as benchmark for our background functional
Eq.~(\ref{eq:eta}).

\subsection{Neighbor-dependent parameters}
\label{subsec:ndp}

As discussed in Section~\ref{sec:model}, the intensities of neighboring
probes can be used to estimate the non-specific binding of a given
probe, because of the design of Affymetrix microarrays.  However,
Eq.~(\ref{eq:eta_local}) takes only five neighbors into account although
each spot on the array is surrounded by in total eight neighbors ---
four direct and four diagonal neighbors.  Eq.~(\ref{eq:eta_local})
originally included eight parameters but it turned out that the
intensity correlations with the ``top'' neighbors at $(y+1)$ (see
Fig.~\ref{fig:orig_param}) on the background intensity are much smaller
than the $(y-1)$ row. The analysis of the correlation between sequences
which are neighbors in the array explains why the $(y+1)$ neighbors are
less useful.

For the HGU133A array all four nucleotides are roughly equally
present, i.e.~A, C, G, T densities are $0.239$, $0.248$, $0.243$ and
$0.269$. A consequence is that with two randomly chosen nucleotides,
the probability of finding the same letter is $25.05\%$. However, the
probability of finding the same letter at the $k$-th position at sites
in $(x,y)$ and $(x+1,y)$ (or equally at $(x-1,y)$) is $48.29\%$. This
probability increases to $96\%$ when considering the neighbors $(x,y)$
and $(x,y-1)$, for even y, which is not surprising as the probes at
these locations are a pair of PM and MM, which share 24 out of 25
oligonucleotides. The probability of finding the identical nucleotide
at $(x,y+1)$ for $y$ even is $37.58\%$. We thus see that the sequence
correlation along the $x$-direction clearly exceeds the correlation
along the $y$-direction except when considering corresponding PM and MM
probes. Because of this, the three ``top'' neighbors were not considered
any further in order to restrict the computational effort to a minimum
(see Fig.~\ref{fig:orig_param}).

The functional minimization shows another interesting pattern: the
three closest neighbor parameters $p_1$, $p_2$ and $p_5$ are positively
correlated with the background signal, while the diagonal neighbors $p_3$
and $p_4$ show negative correlations
(the correspondence between $p_i$ and positions can be deduced from
Eq.~(\ref{eq:eta_local})).
This result (i.e.~$p_1$, $p_2$, $p_5 >0$ and $p_3$, $p_4 <0$) is found
in all chips analyzed.  Typical average outputs on human genome chips
of the Latin Square experiments are

\begin{equation}
\{p_1, \dots, p_5\} \approx \{0.06, 0.08, -0.04, -0.03, 0.35\}. 
\label{eq:p_numbers}
\end{equation}

The interpretation is as follows: The MM signal is most strongly
correlated with its corresponding PM, as reflected by the magnitude of
$p_5$. The sequence at $(x,y)$ is closely correlated with the two MM
neighbors $(x \pm 1,y)$ (parameters $p_1$ and $p_2$), i.e.~a strong
background intensity at $(x\pm 1,y)$ corresponds to a strong background at
$(x,y)$. However, a strong signal of the MM probes at $(x \pm 1,y)$
may also be caused by the presence of complementary target molecules
at high concentrations in solution. The functional corrects for this
with negative coefficients for the signals at positions $(x \pm 1,y-1)$
(parameters $p_3$ and $p_4$).

\subsection{Sequence dependent parameters}
\label{subsec:sdp}

The nearest-neigbor model is widely used to describe the thermodynamics
of duplex formation of nucleic acids in solution as it yields good
approximations of the sequence dependence of duplex stability (see
e.g.~\cite{bloo00}). It is based on the assumption that the stability of
each base pair depends on the identity and orientation of the adjacent
base pairs.  For a given sequence $s$ of $N$ nucleotides the hybridization
free energy is given by:
\begin{equation}\label{eq:DeltaG}
\Delta G = \sum_{k=1}^{N-1} \sum_{\alpha,\beta} 
\delta^k_{\alpha\beta}(s) \Delta G_{\alpha\beta}
+ \Delta G_{\rm init.}(s_1,s_N),
\end{equation}
where $\Delta G_{\alpha\beta}$ are the stacking free energies associated
to a pair of nucleotides $\alpha \beta$; $\delta^k_{\alpha\beta}$ counts
the frequency of the pairs $\alpha \beta$ along the sequence and was
defined in Eq.~(\ref{eq:delta_alphabeta}). In Eq.~(\ref{eq:DeltaG}) we
have added a term which depends on the end nucleotides $s_1$ and $s_N$
and it is referred to as helix initiation parameter $\Delta G_{\rm init}$.

The parameters  $\Delta H_{\alpha\beta}$ and $\Delta S_{\alpha\beta}$
from which one obtains $\Delta G = \Delta H - T \Delta S$ are known from
hybridization experiments in solution. Due to symmetry considerations,
there are only $10$ independent $\Delta G_{\alpha\beta}$ in the case
of DNA/DNA duplexes (see Table 2 of~\cite{sant98}).  There are no such
symmetries in RNA/DNA duplexes hence there are in total $16$ parameters,
which were determined experimentally by Sugimoto et al.~\cite{sugi95_sh}.
Even though the nearest-neighbor model was originally developed to
calculate duplex free energies in solution, it provides reasonable
approximations to describe the energetics involved in the hybridization
processes on Affymetrix microarrays~\cite{held03,carl06}. A recent
experimental study \cite{weck07} on a class of spotted arrays in which
hybridization of perfect matching and multiple mismatching probes were
analyzed, showed that the data are well described by nearest-neighbor
parameters for duplex formation in solution.

\begin{figure}[t]
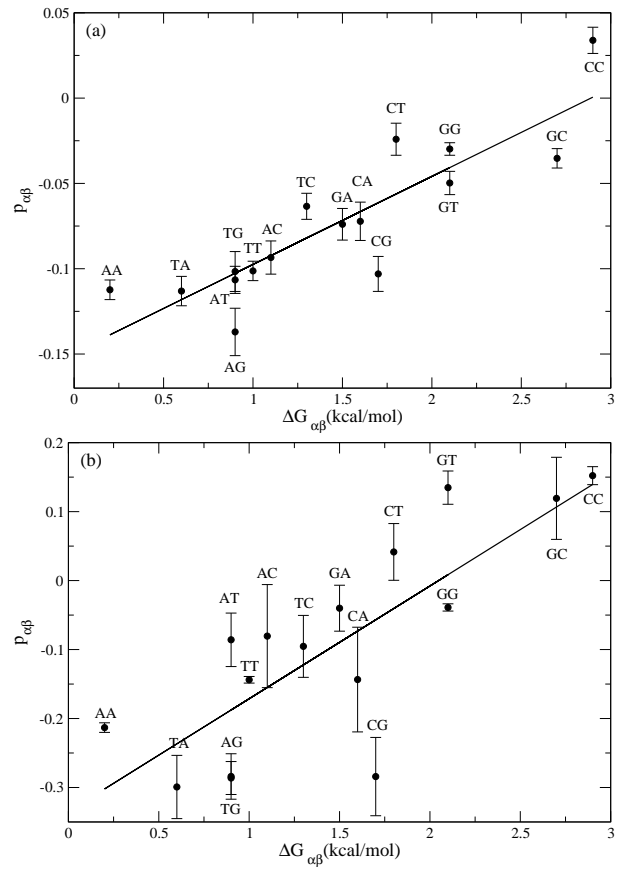
 
\includegraphics[width=8cm]{pXpairssugistdv_out12_hgu95a_1521_resca_pub.eps}
\includegraphics[width=8cm]{pXpairssugistdv_out12_hgu133a_resca_pub.eps}
\caption{Parameters $p_{\alpha\beta}$, as obtained from the minimization
of Eq.~(\ref{eq:costfct}) on a training data set, plotted as function
of $\Delta G_{\alpha\beta}$, the nearest-neighbor stacking free energy
obtained from DNA/RNA hybridization in solution \cite{sugi95_sh}. The
two figures refer to (a) average of 19 experiments of the HGU95 (1521)
spike in data set (b) average of 42 experiments of the HGU133A spike in
data set.  The error bars are the standard deviation.  Notation of DNA
pairs are from 5' to 3' ends. The straight lines are linear fits to the
points. The correlation coefficients of the fits referring to these and
to other experiments analyzed are given in Table~\ref{table:corrcoeff}.}
\label{fig:sugi_hgu} 
\end{figure} 

Our approach to model the background intensity involves the determination
of 24 parameters of which the 16 parameters $p_{\alpha\beta}$ reflect the
influence of each pair $\alpha\beta$ on the background intensity. The
relationship between the 16 parameters $p_{\alpha\beta}$ and the 16
stacking parameters $\Delta G_{\alpha\beta}$ is quickly derived. According
to the Langmuir model the measured intensity $I$ at a given site is
related to the hybridization free energy $\Delta G$ via

\begin{equation}\label{eq:I_prop_expDG}
 I \propto e^{-\Delta G/RT}.
\end{equation} 

Recalling that the sequence dependent functional $\eta_{seq}$ given in
Eq.~(\ref{eq:eta_seq}) is fitted to the logarithm of the intensity (see
Eq.~(\ref{eq:eta})) one expects that the parameters $p_{\alpha\beta}$
are linearly related to the the stacking free energy parameters
$\Delta G_{\alpha\beta}$.  To verify inhowfar this linear relationship
holds, all 16 parameters $p_{\alpha\beta}$ are calculated for each
CEL-file, i.e.~each chip by the minimization of the cost function
Eq.~(\ref{eq:costfct}). Then, each $p_{\alpha\beta}$ is averaged over
all available CEL-files of a given chipset and plotted as a function of
$\Delta G_{\alpha\beta}$ given in Ref.~\cite{sugi95_sh}.  Two of these
plots for the Latin square set are shown in Fig.~\ref{fig:sugi_hgu}.
The plots indicate that the linear relationship between $p_{\alpha\beta}$
and $\Delta G_{\alpha\beta}$ is approximately verified. The correlation
coefficients for the linear fit are typically about $0.83$. The Table
\ref{table:corrcoeff} reports the correlation coefficients for {\it
H. Sapiens}, {\it X. Laevis} and {\it D. Rerio} chipsets.  The results
show that our ansatz to include the nearest-neighbor model in the
background estimation is justified and the influence of the pairs is
not to be neglected.

\begin{table}
\begin{center}
\begin{tabular}{| l | c c| }
\hline
Chipset & \# of CEL-files & Corr. coeff \\
\hline
HGU95A (1521) & 19 & 0.870 \\
HGU95A (1532) & 19 & 0.869 \\
HGU95A (2353) & 19 & 0.861 \\
HGU133A & 42 & 0.791 \\
XL (GSE 3334) & 6 & 0.805 \\
XL (GSE 3368) & 20 & 0.806 \\
XL (GSE 4448) & 31 & 0.792 \\
DR (GSE 5048) & 6 &  0.847 \\
\hline
\end{tabular}
\end{center} 
\caption{Correlation coefficients of the linear fits of the
$p_{\alpha\beta}$ parameters obtained from minimization of the
background functional and the hybridization free energies $\Delta
G_{\alpha\beta}$ for RNA/DNA duplex formation in aqueous solution taken
from Ref.~\cite{sugi95_sh}.  The data are for human chipsets (HGU95A
sets, HGU133A) of the Affymetrix Latin square experiment and of {\it
Xenopus Laevis} (XL) and {\it Danio Rerio} (DR) arrays.}
\label{table:corrcoeff}
\end{table} 

\subsection{Benchmark: Spike In Data}
\label{subsec:bSId}

To test the accuracy of the predicted background signal as
given in Eq.~(\ref{eq:eta}), we turn our attention to the spike-in
data. Concerning background analysis, we are naturally most interested
in the $c=0$ spike-in data as the measured signal is pure background
noise. By virtue of Eq.~(\ref{eq:eta}) we calculate the background signal
$\eta$ for a given probe set of the Latin Square data for the chipsets
HGU95A and HGU133A.

\begin{figure}[t]
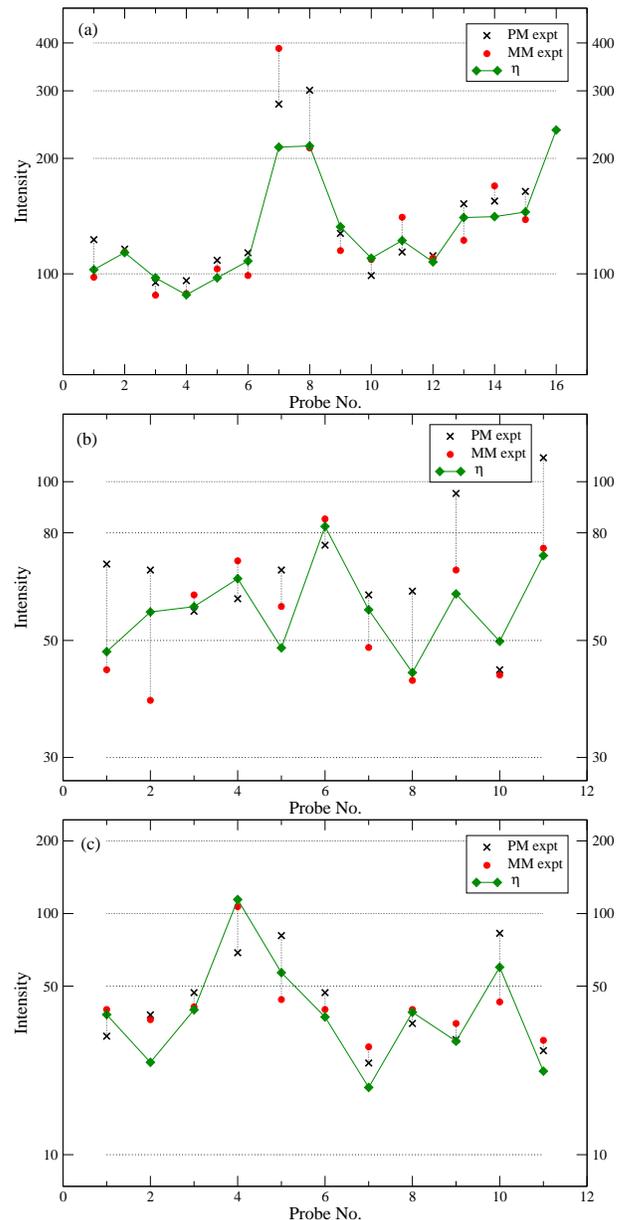

\includegraphics[width=8cm]{38734_at_out12_1521l_pub.eps}
\includegraphics[width=8cm]{AFFX-r2-TagE_at_out12_U133A_Expt4_R1_pub.eps}
\includegraphics[width=8cm]{209795_at_out12_U133A_Expt13_R1_pub.eps}
\caption{Signal intensities for PM (crosses) and MM (circles) for three
probe sets plotted as function of the probe numbers. The data are for
three spikes at $c=0$ (zero concentration means that the target are
absent from the solution). The probe sets are (a) 38734\_at (HGU95A -
1521l), (b) AFFX-r2-TagE\_at (HGU133A - Expt4\_R1) and (c) 209795\_at
(HGU133A - Expt13\_R1). The solid line shows the background estimate
based on the functional of Eq.~(\ref{eq:eta}).}
\label{fig:pmmmeta}
\end{figure}

Fig.~\ref{fig:pmmmeta} is representative for the results of HGU95A
and HGU133A. In general, we find that the predicted background
intensity $\eta$ nicely follows the PM/MM intensities of the spike-in
experiments at zero concentration and hence really describes the
shape of the background. One would expect the PM and MM values at zero
concentration to be almost identical; this is mostly the case as the
median value of the difference $\text{(PM-MM)}_{\text{HGU95A}}=28$
for HGU95A shows. This value is even smaller for the HGU133A chipset,
i.e.~$\text{(PM-MM)}_{\text{HGU133A}}=21$. Exceptions where either
the PM or MM intensity clearly exceeds the median difference suggest
the presence of transcript fragments which are complementary to the
probe over a length of more nucleotides than one would statistically
expect when considering background issues. Especially the origin of
bright MM's has been investigated intensively in the recent past (see
e.g.~\cite{naef03,carl06b,naef06}).

\begin{figure}[t]
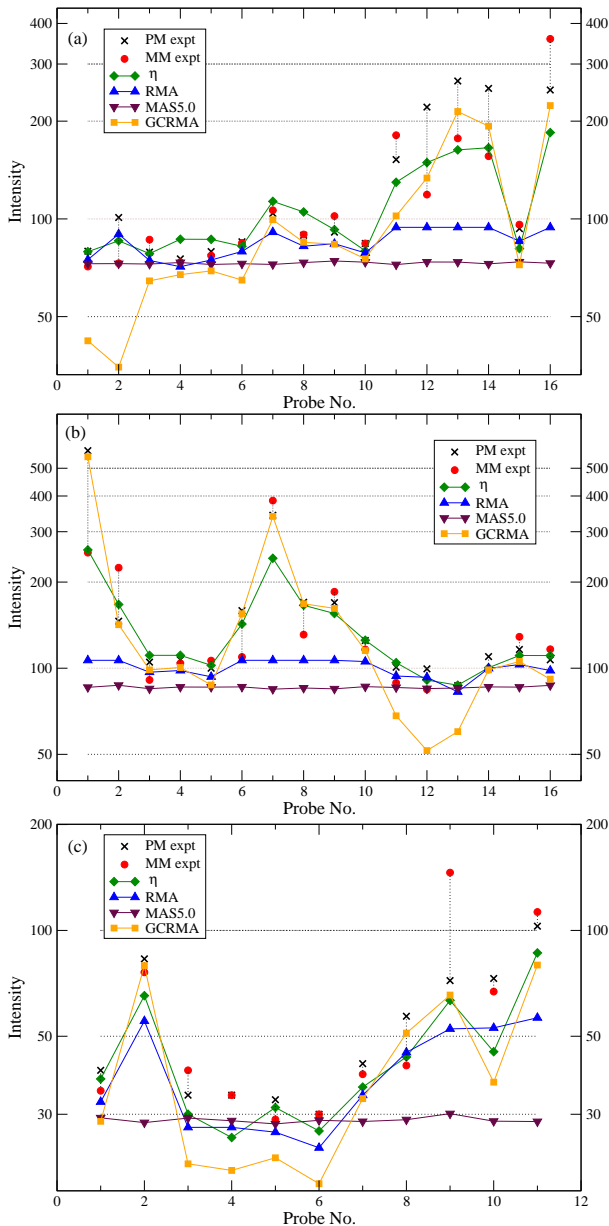

\includegraphics[width=8cm]{36202_at_RMAMASGCRMAout12_1521g_pub.eps}
\includegraphics[width=8cm]{1708_at_RMAMASGCRMAout12_1532b_pub.eps}
\includegraphics[width=8cm]{209606_at_RMAMASGCRMAout12_U133A_Expt10_R1_pub.eps}
\caption{Examples of comparison of the performance of MAS5.0
(triangles down), RMA (triangles up) and GCRMA (squares) with
the algorithm developed in this paper (diamonds). The crosses (PM) and
circles (MM) are the zero concentration spike-in data.  The data shown
are for the probe sets (a) 36202\_at (HGU95A - 1521g), (b) 1708\_at
(HGU95A - 1532b) and (c)209606\_at (HGU133A - Expt10\_R1).}
\label{fig:other_comp}
\end{figure} 

\subsection{Comparison to other approaches}

Figure~\ref{fig:other_comp} compares the performance of our background
functional $\eta$ to three of the most commonly used algorithms,
namely MAS5.0~\cite{affy_statalgo,affy_mas5}, RMA~\cite{iriz03} and
GCRMA~\cite{wu05}. MAS5.0 is a commercial software for data analysis
developed by Affymetrix.  For our calculations we used the free version
of MAS5.0 available under the open project Bioconductor \cite{gent03}.
RMA and GCRMA are two variants of the same type of algorithm, both freely
available from Bioconductor.

In order to compare the performance of the background subtraction 
schemes, we calculated
\begin{equation}
d = \frac{1}{M}\,\sum [\log{I_{PM}} - \log I_b]^{2},
\label{eq:avges}
\end{equation} 
i.e.~the average squared deviation of the predicted background signal $I_b$
from MAS5.0, RMA, GCRMA and from our algorithm
with respect to the experimental background intensity $I_{PM}$. 
The sum in Eq.~(\ref{eq:avges}) runs over all $M$ probes 
in the Affymetrix spike-in experiments at concentration $c=0$.

The examples of Fig.~\ref{fig:other_comp} show that MAS5.0
underestimates the background values and hardly deviates from a
straight line. MAS5.0 uses the lowest 2\% of probe intensities of
each region of a chip to estimate a background value. Each probe
intensity is then background corrected based upon a weighted average
of each of the background values. A detailed description can be found
in~\cite{affy_statalgo,affy_mas5}.  The background adjustment method
used by RMA~\cite{iriz03} uses a global model for the distribution of
probe intensities. It is based on empirical findings on the distribution
of probe intensities and only considers PM values as input as well as
output. However, RMA does not take non-specific binding into account which
often leads to an underestimation of the background.  GCRMA~\cite{wu05}
is based on RMA and includes sequence information to calculate a
so-called affinity measure. The results of GCRMA excel those of RMA
and MAS5.0. However, we have found that in some cases after background
subtraction GCRMA gives a higher value of the intensity compared to the
original data, which signifies a {\it negative} background correction. For
these points we have set $\log I_b = 1$ in Eq.~(\ref{eq:avges}).

Table~\ref{table:avges} reports the value of the mean squared deviation
calculated from Eq.~(\ref{eq:avges}). Smaller values of this parameter
signify a more accurate algorithm for the background estimation.
The Table indeed shows that globally our physical-chemistry based
algorithm, indicated as column $\eta$, outperforms the three
other statistical-based algorithms. As already anticipated by the graphs
in Fig.~\ref{fig:other_comp}, the performance of GCRMA is generally
far better than MAS5.0 and RMA. Our algorithm improves further on GCRMA
in all cases analyzed, except for the last set (HGU95A expertiment 2353)
of Table \ref{table:avges}. 

\begin{table}[h!]
\begin{center}
\begin{tabular}{| l | c  | c c c|}
\hline
$d$ & $\eta$ & RMA & MAS5.0 & GCRMA\\
\hline 
HGU133A     & 0.161 & 0.521 & 1.589 & 0.194\\ 
HGU95A-1521 & 0.163 & 0.760 & 1.127 & 0.200\\ 
HGU95A-1532 & 0.203 & 0.698 & 1.041 & 0.343\\
HGU95A-2353 & 0.099 & 0.508 & 0.777 & 0.088\\
\hline
\end{tabular}
\end{center}
\caption{Average squared deviation of four human genome chipsets according
to Eq.~(\ref{eq:avges}) where $I=I_{PM}$.}
\label{table:avges}
\end{table}

\section{Discussion}
\label{sec:discussion}

We have introduced a new model to predict background intensities in
Affymetrix GeneChips. Our model takes into account the physical-chemistry
involved in hybridization as well as the influence of the design
of Affymetrix microarrays.  The background functional developed in
this paper contains two terms given by Eq.~(\ref{eq:eta_local}) and
Eq.~(\ref{eq:eta_seq}) that reflect these two contributions.

The sequence-based background estimate (Eq.~(\ref{eq:eta_seq})) includes
16 pair-strength-parameters $p_{\alpha\beta}$. Physical-chemistry
arguments suggest that these parameters are correlated to the
hybridization free energies $\Delta G_{\alpha\beta}$ for the corresponding
couple of nucleotides. One expects an approximate linear relationship
between the two. The fact that the parameters $p_{\alpha\beta}$
are indeed linearly correlated to the hybridization free energies
in solution, as shown in Fig.~\ref{fig:sugi_hgu}, suggests that the
model presented here captures the origin of the background correctly.
We recall that hybridization in Affymetrix expression arrays is between
a DNA strand at the microarray surface and an RNA strand in solution,
therefore the hybridization free energies to compare with are those
for RNA/DNA duplexes. These were determined experimentally by Sugimoto
et al.~\cite{sugi95_sh}.  It is worth mentioning that a previous study
\cite{zhan03} of microarray data analysis using physical-chemistry inputs,
although in a different way than what is depeloped here, reported a weaker
correlation ($r=0.6$) between fitted affinities and the experimental
parameters by Sugimoto et al.~\cite{sugi95_sh}. In the experimental data
considered in this study we find a correlation coefficient ranging from
$r=0.79$ to $r=0.87$ (see Table \ref{table:corrcoeff}) for the three
different organisms analyzed.  In our opinion, a good correlation with
experimental stacking free energies provides a first important test of
reliability of the analysis.

In our model, a second contribution to the background functional is
given by the intensities at the locations that are physical neighbors
on the microarray (Eq.~(\ref{eq:eta_local})). The neighbors influence is
understood as coming from the fact that neighboring locations have similar
sequences, as a consequence of the design of Affymetrix microarrays:
similar sequences imply similar background contributions.  The local
contribution to the background depends on five parameters which measure
the strength of the correlations. As pointed out in Sec.~\ref{subsec:ndp}
the magnitude and signs of these parameters can be understood in terms
of sequences similarities.

We compared the background intensities predicted by the functional presented in
this paper with the experimental data. The latter are spike-in Affymetrix
data \cite{latinsq} in which few sequences are added in solution at
known concentration. The spike-in data set is used to develop and test
algorithms for Affymetrix microarrays data analysis. In particular we
considered the data at zero spike-in concentration, which measure pure background.
We used these data to compare the performance of our algorithm to
the other algorithms MAS5.0, RMA and GCRMA. This comparison is summarized
in Table \ref{table:avges}, showing the average squared deviation from
the logarithm of the intensities at zero spike-in concentration.  The results
show that our algorithm and GCRMA perform much better than both MAS5.0
and RMA. In the tests performed we noticed that GCRMA follows
closely the experimental background, but it may ``fail" substantially
in few probes of a probe set. This can also be seen in the examples
of Fig.~\ref{fig:other_comp}.  These failures lower the performance of
GCRMA, compared to the physical-chemistry algorithm presented here.

In conclusion, the algorithm developed in this paper provides good quality
results for background estimates compared to existing algorithms and
provides an interesting alternative for background subtraction schemes
in Affymetrix Genechips.  Even though we have shown that the performance
of our background functional is satisfying, hopefully there is still
room for improvement.


{\bf Acknowledgment}. We acknowledge financial support from the Van Gogh
Programme d’Actions Int\'egr\'ees 08505PB of the French Ministry of
Foreign Affairs and Grant No. 62403735 and by the Netherlands Organization
for Scientific Research (NWO).  Support from Fonds voor Wetenschappelijk
Onderzoek (FWO) Grant No. G.0297.06 is also gratefully acknowledged.


\end{document}